\documentclass[pre,aps,twocolumn,showpacs]{revtex4}
\usepackage{bm}
\usepackage[dvips]{graphicx}
\usepackage{amsmath}

\begin{document}

\title{How to Track Protists in Three Dimensions}

\author{Knut Drescher, Kyriacos C. Leptos, and 
Raymond E. Goldstein\email{R.E.Goldstein@damtp.cam.ac.uk}}
\affiliation{Department of Applied Mathematics and Theoretical
Physics, Centre for Mathematical Sciences, University of
Cambridge, Wilberforce Road, Cambridge CB3 0WA, UK}


\begin{abstract}
We present an apparatus optimized for tracking swimming microorganisms in the size range 
$10-1000$ $\mu$m, in three dimensions (3D), far from surfaces,
and with negligible background convective fluid motion.  CCD cameras attached 
to two long working distance microscopes synchronously image the sample from 
two perpendicular directions, with narrowband dark-field or bright-field 
illumination chosen to avoid triggering a phototactic response. The images from the 
two cameras can be combined to yield 3D tracks of the organism. Using additional, 
highly directional broad-spectrum illumination with millisecond timing control 
the phototactic trajectories in 3D of organisms ranging from 
{\it Chlamydomonas} to {\it Volvox} can be studied in detail.  Surface-mediated
hydrodynamic interactions can also be investigated without convective interference.
Minimal modifications to the apparatus allow for studies of chemotaxis and
other taxes.
\end{abstract}


\maketitle

\section{Introduction}

\begin{figure}[b]
\begin{center}
\includegraphics*[clip=true,width=0.95\columnwidth]{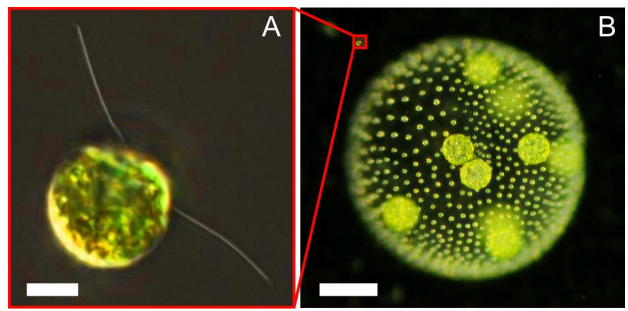}
\end{center}
\caption{\label{fig:protists} Two protists whose swimming motion is of interest 
in this work.  (A) {\it Chlamydomonas reinhardtii} (scale bar $5$ $\mu$m) and 
(B) {\it Volvox carteri} (scale bar $200$ $\mu$m).}
\end{figure}

Protists, the grouping of eukaryotic microorganisms that encompasses such diverse entities 
as flagellated and ciliated protozoa \cite{protistology} (e.g., {\it Euglena}, 
{\it Paramecium}) and motile green alga \cite{greenalgae}  (e.g., {\it Chlamydomonas}, 
{\it Volvox} -- shown in Fig. \ref{fig:protists}), constitute an important class of organisms 
in the study of evolutionary biology, biological physics, and, recently, biological fluid 
dynamics \cite{multicellular,flagflows,Takuji_paramecium}. Many flagellated protists display 
swimming behavior that is inherently three-dimensional (3D). A number of important 
questions in biology and physics are associated with how the motion of such organisms 
is related 
to their body plan and to external stimuli such as light \cite{Hill_phototaxis,Schaller}, 
dissolved molecular species, gravity, temperature, boundaries \cite{waltzing}, and 
electromagnetic fields. It is thus desirable to track their position and orientation 
in 3D \cite{Vlad1,Vlad2} with high spatio-temporal resolution and, unless desired, 
free from systematic bias introduced by external stimuli, background fluid motion, and 
hydrodynamic surface effects \cite{surfeffect,squires2001}.

The first apparatus able to track microorganisms in 3D was designed for bacteria 
\cite{berg1971} and utilized an analogue feedback loop that moved the microscope 
stage to keep a bacterium centered in the field of 
view.  Larger microorganisms such as protists require larger sample chambers, reaching 
millimeters or even centimeters in depth to avoid boundary-induced hydrodynamic effects.  
In this regime, methods based on a moving microscope stage are not suitable, as they 
induce uncontrolled background fluid motion in the sample chamber. 
Similar considerations enter the study of the millimetric
nematode {\it C. elegans} crawling on the surface of agar, 
for which a moving substrate introduces unwanted mechanical stimuli.  Instead,
the camera itself can by moved by motors controlled by an algorithm 
that dynamically centers the worm in the field of view \cite{ryu}.
Several apparati for 
tracking the 3D motion of microscopic particles without a moving stage have emerged in 
recent years, yet none is ideal for studying the swimming of protists.  Methods 
based on controlled defocussing of particles \cite{wu2005, willert1992}, placing a 
cylindrical lens in the imaging optics of the microscope \cite{kao1994}, or measuring 
the deflections of a laser beam that is focused close to a particle 
\cite{peters1998,ghislain1994} all suffer from a small tracking range along the optical 
axis. Observing a sample which is illuminated from the side with a continuous 
gradient of color
in order to color-code the third dimension 
\cite{matsushita2004, mcgregor2008} suffers from low spatial resolution in that dimension, 
and may provide a photostimulus to protists. Tracking objects with a confocal microscope is 
only possible when the objects move at very low speeds \cite{dinsmore2001,rabut2004}. Digital 
in-line holography may also be used for 3D particle tracking, yet even vibrations 
with amplitudes $<1$ $\mu$m of components along the optical path lead to a 
time-varying background in the hologram that can significantly degrade the signal of 
the moving object.

The difficulties mentioned above can be overcome by using more than one camera to observe 
synchronously the sample chamber from different angles
and then combining the images to yield 3D tracks of particles.
Yet, existing implementations of this technique can not be 
easily adapted to track microscopic objects \cite{maas1993,hoyer2005}; they 
either have a  
controlled but undesirable shear flow \cite{grillet2007}, have not integrated any 
control of thermal convection \cite{strickler1998}, or have been implemented with a 
temperature gradient across the sample ($\sim 3$ K/cm) to eliminate thermal 
convection \cite{thar2000}.  The latter method is an important advance, but 
may introduce an unwanted behavioral stimulus to protists. 

Here we present an apparatus that uses two identical imaging assemblies at 
right angles to each 
other that can be used in dark- and bright-field illumination, combined with systems to 
deliver photo-stimuli, to control temperature inside the sample, and to eliminate background 
fluid motion in large sample chambers (up to $2.5 \times 2.5 \times 5$ cm) filled with 
aqueous growth medium, by homogenizing the temperature inside the sample chamber 
to millikelvin precision. The apparatus uses almost entirely off-the-shelf components 
and requires minimal expertise in optics. Online supporting material 
includes software to control the hardware and to perform 3D tracking.  We 
discuss the resolution and limitations of the apparatus, present swimming 
trajectories of the protists {\it Chlamydomonas} and {\it Volvox} in 3D, and illustrate 
with dual-view particle imaging velocimetry (PIV) the flow field that {\it Volvox} 
generates near a surface.

\section{Experimental Apparatus}
\label{sec:experimentalapp}

\begin{figure}[t]
\begin{center}
\includegraphics*[clip=true,width=0.90\columnwidth]{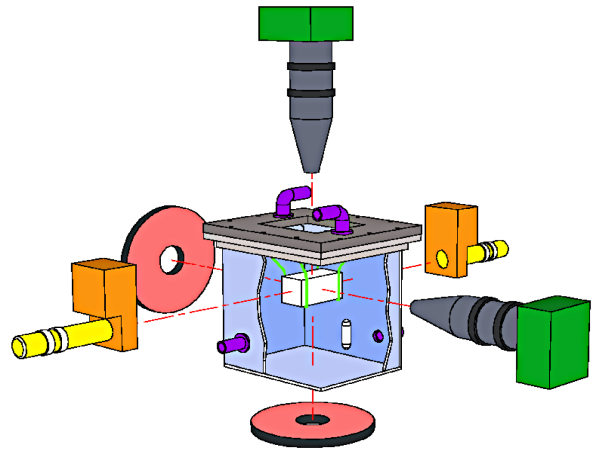}
\end{center}
\caption{\label{fig:apparatus} Schematic drawing of apparatus. 
The outer chamber (light blue, lid and flange in grey) contains a water bath, 
fed through inlets and outlets (purple), and mixed with a magnetic stir bar (white) driven
by a motor external to the tank (not shown). 
The sample chamber (white) is suspended by stainless steel holders (light green), and 
illuminated by annular LED arrays (red).  Microorganisms are visualized with
two long-working-distance microscopes (dark gray) equipped with CCD cameras (dark green). 
Phototactic stimulus is provided by two LED and lens assemblies (yellow), 
and controlled by shutters (orange).}
\end{figure}

The 3D tracking system, shown schematically in Fig. \ref{fig:apparatus}, is based on a 
flexible but powerful imaging system, phototactic 
stimulus lights, and equipment to control and homogenize the temperature inside 
the sample chamber. These three elements are now 
explained in detail.

The imaging system is comprised of two identical assemblies that are mounted at 
right angles on a vibration isolation table
(Science Desk, with $900\times 1200\times 60$ mm breadboard, Thorlabs, Ely, UK). 
A monochrome FireWire CCD camera 
(Pike F145B, Allied Vision Technologies, Stadtroda, Germany; 
$1388\times 1038$ pixels, each $6.45\times 6.45$ $\mu$m, 
maximal frame rate of $30$ fps, and support for 
external triggering) was attached to each of the two microscopes
(InfiniVar CFM-2/S, Infinity Photo-Optical, Boulder, CO).  These 
are continuously focusable with a working distance between $18$ mm and $\infty$, yielding 
a maximum magnification of $\times 9$ at the smallest available working distance.
To allow a variable working distance, the camera/microscope assemblies 
were mounted on sliding rails (PRL-12, Newport Corp., Irvine, CA) via standard
post/post support hardware. The horizontal rail was attached directly 
to the breadboard while the vertical one was attached to a movable and lockable 
rail carrier on a large optical 
rail (X95, Newport Corp.) mounted vertically to the
optical breadboard.  The outer chamber 
(see Fig. \ref{fig:apparatus}, and details below) limits the smallest working 
distance to $\sim 60$ mm, yielding a magnification of $\times$1.  While sufficient 
to image protists of size $\sim10$ $\mu$m with dark-field illumination,
such organisms are better visualized with an 
additional $\times2$ magnifier lens (2xDL, Infinity Photo-Optical). This 
increases the working distance and the depth of field, as discussed below. 

The imaging system can be used with dark- or 
bright-field illumination: bright-field can be advantageous for organisms larger 
than $\sim 100$ $\mu$m as it captures more details of the organism ({\it e.g.}, the 
body axis); dark-field can be desirable when the organism is so small that only 
center-tracking is possible, as under these conditions it yields a better 
signal-to-noise ratio. The flexibility to use both illumination techniques is obtained 
by using an annular LED array (LFR-100-R, CCS Inc., Kyoto, Japan) as the (unpolarized) 
light source 
for each microscope. Dark-field illumination is achieved when the field of view of the 
microscope only includes the dark region in the center of the LED annulus. 
Bright-field illumination is achieved by inserting a diffuser plate (bulk frosted acrylic 
cut to size, RS components, UK) in between the LED array and the outer chamber, 
as far from the LED array as possible. We chose the color of the LED array to be narrowband 
red ($655$ nm, $21$ nm bandwidth) as it has been shown that this color does not trigger a 
phototactic response in motile green algae \cite{volvox_actionspectrum1}. 

For phototaxis studies, two opposing light sources are required in order to observe 
reproducible light-induced U-turns. The photo-stimulus lights were two broad 
spectrum cool white Luxeon LEDs (MWLED, Thorlabs), collimated with a 
short focal length lens ($f=35$ mm, LA1027, Thorlabs), and mounted to a shutter with millisecond 
precision (SH05, Thorlabs). The shutter/LED assemblies are mounted via standard
cage hardware (Thorlabs) to posts on rotating stages (RP01, Thorlabs), allowing the
direction of the light to be controlled. The stimulus direction was typically 
chosen to be horizontal in order to avoid an additional gravitational stimulus. This 
choice forces the common axis of the two cameras to be horizontal, along the stimulus 
direction. The beam diameter was controlled with an iris (SM1D12, Thorlabs) to 
illuminate only those faces of the sample chamber perpendicular to 
the common axis of the two cameras, thereby avoiding reflections,
and the resulting unclear stimuli, from the other four faces.

The two CCD cameras and shutters for the photo-stimulus lights were controlled 
with {\scshape labview} including the add-on toolbox {\scshape ni-imaq} 
for {\scshape 1394-ieee} cameras (National Instruments, Austin, TX), allowing 
precise synchronization of image acquisition. A {\scshape labview} 
program is part of the supporting online material.

In order to combine the images from both cameras to yield 3D 
swimming tracks, it is crucial that both microscopes operate at the 
same magnification (i.e. the same working distance). This is easily achieved 
to sufficient precision by replacing the sample chamber with a tilted calibration 
ruler that is 
observable through both cameras, and adjusting the 
working distance until the field of view of both cameras has the same physical size. 
After this calibration step, the microscopes need to be aligned along their common 
axis such that the field of view of both cameras contains the same section of the 
common axis. This axis alignment greatly improves the ability to reconstruct 3D tracks, 
as explained in Section \ref{sec:improcess}.

To obtain swimming trajectories that are not influenced by hydrodynamic surface 
effects or background flows, it is desirable to have a sample chamber that is as 
large as possible, while maintaining the fluid within it perfectly still. 
A stationary fluid can only be obtained if the temperature in the chamber and of
the chamber walls is very homogeneous, thereby eliminating thermal convection caused by 
heating from the two LED arrays (each LED array consumes $\sim 3.6$ W). 
For a closed chamber, the critical Rayleigh number above which thermal convection starts 
to occur is $Ra_{\rm c} = \alpha g L^3 \Delta T/\nu \kappa \simeq 1708$,
where $\alpha$ is the thermal expansion coefficient of the fluid, 
$g$ is the gravitational acceleration, $\kappa$ is the thermal diffusivity, 
$\nu$ is the kinematic viscosity, and $L$ is the length scale across which there is 
a temperature difference $\Delta T$ \cite{criticalRayleigh}.  While the precise value 
of $Ra_{\rm c}$ depends on the geometry of forcing and on boundary conditions, the
scale of temperature differences involved for water 
($\alpha=2\times 10^{-4}$ K$^{-1}$, $\kappa=1.4\times 10^{-3}$ cm$^2$/s, 
$\nu=0.01$ cm$^2$/s) is roughly $100$ mK for a one centimeter length. 
The largest chambers that have previously been temperature-homogenized below the thermal 
convection threshold, under comparable conditions to those presented here, have $L \sim 1$ cm in 
sedimentation studies \cite{segre1997,weitzNbody}. 
The system presented here eliminates thermal convection in chambers as large as 
$2.5 \times 2.5 \times 5$ cm, implying temperature differences 
between faces of the chamber to be below $\sim 8$ mK.  

\begin{figure}[t]
\begin{center}
\includegraphics*[clip=true,width=0.90\columnwidth]{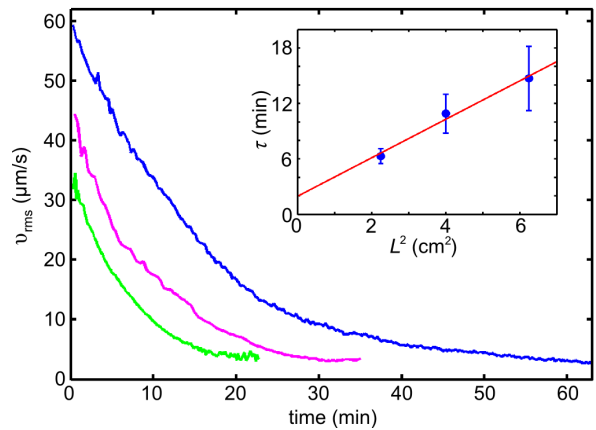}
\end{center}
\caption{\label{fig:convection} Decay of convective motion in the inner sample chamber.
Curves show the root-mean-squared velocity $v_{\rm rms}$ within the chamber, 
obtained by PIV, as a
function of time for three different chamber dimensions: blue - $2.5\times 2.5\times 2.5$ cm,
magenta - $2.0\times 2.0\times 2.0$ cm, green - $1.5\times 1.5\times 1.5$ cm. 
Inset shows the 
time constant for each decay as a function of chamber size, consistent with the $L^2$ 
diffusive scaling.}
\end{figure}

Recognizing that in Stokes flow the effects of boundaries at 
a distance $h$ from compact objects acted upon by gravity fall off as $h^{-1}$, 
and given that swimming trajectories can easily sample a vertical scale that is 
$5$ times the organism diameter, the sample chamber should be $>20$ times the organism 
diameter for surface effects to remain below the 5\% level \cite{surfeffect,squires2001}.  
A chamber of this size thus allows protists as large as $1000$ $\mu$m to be studied with 
negligible hydrodynamic surface effects.

Before explaining how the temperature in the sample chamber is controlled and 
homogenized, it is necessary to give details of the outer and sample chambers. 
The outer chamber has dimensions $12 \times 12 \times 10$ cm, two inlets and 
two outlets (as shown in Fig. \ref{fig:apparatus}) and is made from $2.75$ mm 
thick borosilicate glass (custom made by Fine Glass Finishers Ltd., Great Chesterford, UK). 
The flange and lid of the outer chamber were custom made out of PVC with a CNC machine. 
The three types of sample chamber used were (i) custom cuvettes made 
from standard microscope slides cut with a tungsten glass scriber 
(LAC-450-A, Fisher Scientific, UK) and glued together with UV-curing optical 
glue (NOA68, Norland Products, Cranbury, NJ), cured in a UV Chamber (ELC-500,
Electro-Lite Corporation, Bethel, CT), 
(ii) non-standard commercial glass cuvettes (VitroCom, Mountain Lakes, NJ), and 
(iii) standard glass cuvettes. The sample chamber is held rigidly in the center of the outer 
chamber by four thin stainless steel rods ($1.6$ mm diameter). Each rod was inserted into a
corresponding tight-fitting hole in the lid and fixed with a set screw,
allowing easy modification of the holding arrangements for the different chambers.

It is well-known that the flagella of protists such as {\it Chlamydomonas} have a
strong tendency to stick to glass surfaces \cite{flagella_sticking}.  As cells
swim around the chamber they inevitably collide with the chamber walls.  To avoid
any problem with sticking, the glass is coated with PDMS (Sylgard $184$, Dow-Corning, Belgium) 
and etched for $3$ minutes in a plasma cleaner (Femto, Diener Electronic, Nagold, Germany), 
following a published protocol \cite{whitesides}.
   
The temperature in the sample chamber is controlled by cycling filtered water 
through the outer chamber from a large tank that contains a submersed 
thermostatic heater (J\"ager Eheim, Deizisau, Germany) and pump 
(Eheim, Deizisau, Germany).  In order to homogenize the temperature in the sample 
chamber, the pump is switched off, and strong rare earth magnets 
(EP200, e-Magnets UK) are spun at $\sim 300$ rpm by a sturdy motor 
(178-5112, RS components, UK) on the outside of the outer chamber, thereby 
moving a Teflon-coated $5$ cm magnetic stir bar inside the outer chamber at 
the same speed. This stirring evens out the temperature within the outer chamber 
and, if the stir bar is spun at the appropriate speed and place in the outer 
chamber, moves the water past the faces of the sample chamber without setting 
up recirculating vortices on the faces. By injecting inexpensive tracer 
particles (size $\leq$75 $\mu$m, Pliolite VTAC-L, Eliokem, Villejust, France) 
into the outer chamber in order to visualize the flow across the faces 
of the sample chamber, we found that the arrangement drawn in Fig. 
\ref{fig:apparatus} can homogenize the temperature inside the sample chamber 
below the threshold $Ra_{\rm c}$.  

The flows within the sample chamber were quantified with commercial PIV software 
(FlowManager, Dantec Dynamics, Skovlunde, Denmark).  As a metric for the extent of
flows throughout the chamber, we report the r.m.s. velocity $v_{\rm rms}$ obtained by uniform
averaging over the field of view of one camera.   
Figure \ref{fig:convection} shows that $v_{\rm rms}$ decays
exponentially as a function of time after the stirring of the water bath has
begun, with a time constant $\tau(L)$ that depends on the smaller chamber dimension 
$L$.  It is straightforward
to see when the temperature difference falls below the threshold for
convection, as tracer particles that are used to visualize the convective flow in 
the sample chamber will suddenly begin 
to fall out of the fluid at approximately their Stokes sedimentation speed, 
forming a sedimentation front that propagates downward. For $10$ $\mu$m 
latex beads (C37259, Invitrogen, Carlsbad, CA), the Stokes sedimentation speed 
is $\sim 3$ $\mu$m/s.  This serves as a lower bound for $v_{\rm rms}$ in 
Fig. \ref{fig:convection}. 
We expect $\tau(L)$ to arise from 
viscous dissipation, and thus to scale as $\tau(L)\sim L^2/\nu$.  In water this yields 
times on the order of a few to ten minutes for $L$ in the range $1-2.5$ cm, consistent with 
the data in Fig. \ref{fig:convection}.   Each curve conforms well to a single
exponential decay, and the fitted times $\tau(L)$ 
obey well the expected quadratic scaling as shown in the inset to Fig. \ref{fig:convection}, 
with a value not far from that expected from the kinematic viscosity of water.

\section{Tracking Software}
\label{sec:improcess}
The 3D tracking was done by analyzing the image sequences from each camera 
separately, giving a set of two two-dimensional (2D) tracks, and 
combining suitable tracks from these two sets to yield 3D trajectories.

Modified \textsc{matlab} (MathWorks, Natick, MA) versions of freely available 
\cite{eric_weeks_web} particle tracking routines written by J.C. Crocker and D.G. Grier were 
used for the 2D tracking, allowing many organisms to be tracked at the same time.  
To track organisms that are so small that they appear circular and without internal structure 
in dark-field images (e.g., {\it Chlamydomonas}), no modifications to the original versions of 
the code need to be made.  To track extended objects (e.g., {\it Volvox}) the routines that 
identify the center of the object need to be modified in a way that depends on the shape and 
structure of the object. For bright-field images of the spherical {\it Volvox}, all 
internal structure of {\it Volvox} was removed by histogram equalization, followed by spatial 
band pass filtering. The resulting image was then convolved with a binary disk-shaped kernel, 
yielding an image in which the centroids of peaks correspond to the {\it Volvox} centers in 
the original image.  A {\it Volvox} colony carries daughter colonies inside it 
(see Fig. \ref{fig:protists}), which are fixed in the posterior hemisphere and therefore 
act as convenient markers of the body axis. The axis of {\it Volvox} can thus be 
determined by finding the vector between the geometric center of {\it Volvox} and 
the center of brightness of the daughter colonies inside {\it Volvox} for both directions 
and then combining these two vectors to obtain a 3D axis. The modified code also allows 
additional information for each {\it Volvox} to be gathered, such as the orientation of
the body axis. 

To identify two 2D tracks that are suitable for synthesizing into a 3D track, the two sets 
of 2D tracks were compared along the common axis of the field of view of the two cameras.  
Consider one of the possible combinations of two 2D tracks. Even if these two tracks are 
projections of positions of a single organism, the tracks usually do not completely overlap in 
time, because during the course of a long track the signal from the tracked object may 
fall below the tracking threshold so that the object `drops out' of the tracking 
data \cite{xu2008}. This means that only the time-overlapping sections of each track can 
be compared. Because of the precise alignment of the field of view of both cameras 
(see Sec. \ref{sec:experimentalapp}), a decision upon whether the two 2D tracks are from the 
same organism can be made by finding the r.m.s. difference between the position-coordinate 
along the common axis. The two 2D tracks for which this value is minimal 
(and below a certain threshold) are then synthesized into a 3D trajectory.  Code that can 
perform all the operations described above is part of the supporting online material.

\section{Performance}
\label{sec:resolution}
The tracking precision of the apparatus has been determined by the standard 
method \cite{gelles,cheezum} of observing fluctuations in the tracked position 
of particles that are fixed between two cover slips. The precision was tested 
at the minimum working distance the apparatus allows (60 mm, corresponding to 
$\times$1 magnification), for two 
different types of objects. For monodisperse 10 $\mu$m latex beads imaged in 
dark-field illumination, the uncertainty in the position was found to 
be $\leq 1.5$ $\mu$m, if the $\times$2 magnifier 
lens is mounted to the microscope. For objects in the size range 
425-500 $\mu$m ({\it Volvox} fixed with iodine), in bright-field illumination, 
the uncertainty was found to be $\leq 1.3$ $\mu$m without the magnifier lens.

The performance of the apparatus is determined not only by the 
uncertainty in spatial position, but also by the overall volume in which 3D 
tracking can be performed. This volume is set by the depth of field of the 
microscope. For the purpose of simply tracking the center of a spherical object, 
the object can be out of focus as long as the signal in the image intensity profile 
is sufficiently large. Therefore the ``trackable depth" (TD), 
which we define as the depth in which the signal/noise $\leq 4$, is a more 
suitable measure for the trackable volume than the depth of field. The TD 
is strongly dependent on the object size and on the signal (dark-field 
illumination gives a larger signal). We measured the TD for 10 $\mu$m 
beads in dark-field to be $5.8$ mm ($11.9$ mm) at a working distance of 
$60$ mm ($100$ mm). Imaged in bright-field, fixed {\it Volvox} of size 
425-500 $\mu$m have TD $= 18.3$ mm at the minimum 
working distance of $60$ mm.

A limitation of the tracking software presented here is that the concentration of 
organisms in the sample chamber should not be so large that swimmers overlap frequently 
in the 2D images from each camera.  As the tracking software can not distinguish overlapping 
objects, a high concentration of swimmers would result in very short 2D tracks, 
and therefore less accurate synthesizing of 2D tracks to a 3D track. An alternative 
method for obtaining 3D tracks from the images of more than one camera is to determine 
the 3D position of every particle at every time point. This approach is often taken in 
3D Lagrangian particle tracking \cite{maas1993,kieft2002}, and can handle larger 
concentrations of trackable objects, but is usually implemented with at least three 
cameras in order to reduce frequent ambiguities in the 3D particle identification, and 
requires an elaborate calibration. This 3D tracking method could be implemented with the 
apparatus described here, if one of the stimulus lights is replaced by 
a third microscope-camera-assembly.

Another limitation is the size of the outer 
chamber, which limits the magnification to values that are not 
sufficient for tracking small bacteria (e.g., {\it E. coli}), even though 
the microscope has a maximum magnification of $\times$18 (including the 
$\times$2 magnifier lens). For tracking bacteria sample chambers 
of size $2 \times 2 \times 2$ mm \cite{berg1972} may be used, for which
the temperature-homogenizing outer chamber is not needed.

\begin{figure}[t]
\begin{center}
\includegraphics*[clip=true,width=0.9\columnwidth]{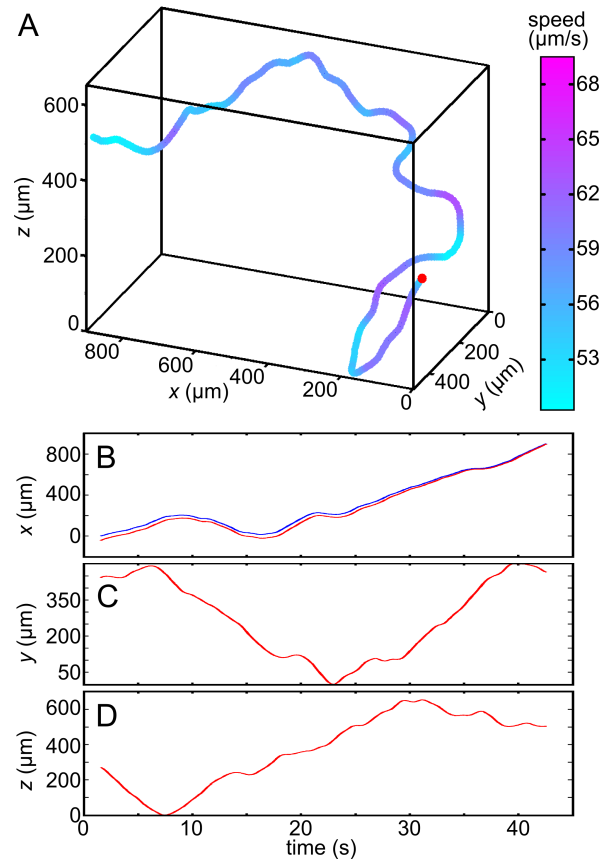}
\end{center}
\caption{\label{fig:reconstruction} Reconstruction of a swimming trajectory of
{\it Chlamydomonas reinhardtii}.  Gravity is along negative $z$-direction.
(A) 3D trajectory, color-coded to indicate local speed. Red dot indicates start of the trajectory. 
(B)-(D) Three components of position versus time.  In (B) are the two traces $x(t)$ from
the two cameras.}
\end{figure}

\section{Application of apparatus to track {\it Chlamydomonas} and {\it Volvox}}
\label{sec:application}
The apparatus was tested on two low-Reynolds number swimmers of very different size: 
{\it Chlamydomonas reinhardtii} (diameter of $\sim 10$ $\mu$m) and 
{\it Volvox barberi} (diameter of $\sim 600$ $\mu$m). Both species 
were grown axenically in Standard {\it Volvox} Medium (SVM) \cite{svm} with sterile air bubbling, 
in diurnal growth chambers (Binder KBW400, Tuttlingen, Germany) set to a daily cycle 
of $16$ h in cool white light ($\sim 4000$ lux) at $28^{\circ}$ C and $8$ h in the 
dark at $26^{\circ}$ C. Sample chambers were filled with SVM 
and were of size $2.5 \times 2.5 \times 5$ cm for {\it Volvox}, and 
$1 \times 1 \times 4$ cm (standard cuvette) for {\it Chlamydomonas}. 

The biflagellated {\it Chlamydomonas} beats its flagella at $\sim 40$ Hz,
primarily in the manner of the breast stroke.  Its most familiar swimming trajectory is
helical, with a radius of $20$ $\mu$m and a speed on the order of $50$ $\mu$m/s.  
The cell has an ``eye-spot" that serves as a photosensor, and the   
changing illumination levels of the eye-spot lead to transient changes in flagellar 
beat dynamics in such a manner that the cell can turn toward the light.  

Figure \ref{fig:reconstruction}A shows a $45$ s trajectory,
obtained without phototactic stimulus, during which the cell explored a volume 
less than $1$ mm$^3$.
The manner in which the two 2D trajectories from the cameras are synthesized into a
3D trajectory is indicated in panels (B)-(D) of the figure.  The $x$-axis 
is common to the two cameras, and the overlap between the two is clear in Fig. 
\ref{fig:reconstruction}B.  The very slight graded mismatch between the two $x$-component
curves reflects a slight misalignment of the cameras, and is easily removed by
remapping the pixel coordinates.

\begin{figure}[t]
\begin{center}
\includegraphics*[clip=true,width=0.9\columnwidth]{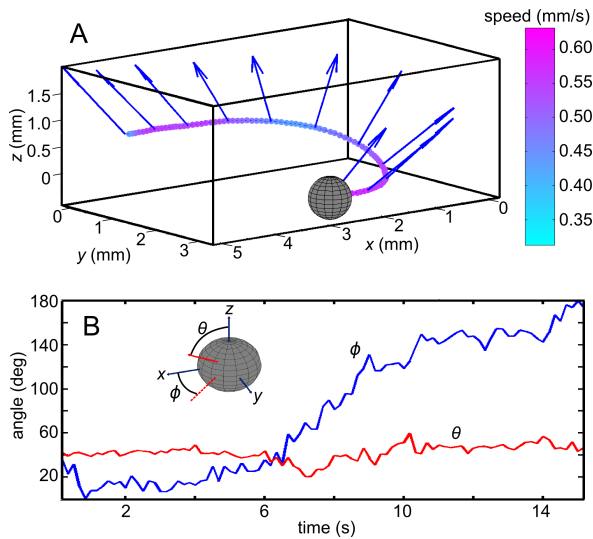}
\end{center}
\caption{\label{fig:phototurns} A phototactic turn of {\it Volvox barberi}.  (A) The
positional and orientational measurements are illustrated 
by vectors indicating the body axis, and swimming-speed-dependent coloration of 
the track. To initiate the 180$^\circ$ change swimming direction, the light initially
was from the right along the $x$-axis and then changed to come from the left.
Gravity is directed along the negative $z$-direction.  Sphere represents initial
position along the track. (B) Evolution of the body axis during the phototactic turn 
is described in terms of two angles $\theta$ and $\phi$.}
\end{figure}

{\it Volvox barberi} typically has $10,000-50,000$ biflagellated somatic cells
rigidly embedded at the surface of a transparent extracellular matrix.  These beat
at a typical frequency of $20$ Hz, primarily from the anterior pole to the posterior 
pole, with a slight tilt of the beat plane which leads to the characteristic spinning motion
as it swims at speeds up to $800$ $\mu$m/s \cite{solari_barberi}.  
Just as in {\it Chlamydomonas}, each
somatic cell has an eye-spot that modulates the beating of its two flagella, 
allowing the whole colony to perform phototaxis.
A 3D track of the {\it Volvox} center and body axis during a phototactic turn is shown 
in Fig. \ref{fig:phototurns}A. Determining the body axis of {\it Volvox} by using the 
position of the daughter colonies, as explained in Sec. \ref{sec:improcess}, leads to 
the time series of the angles $\theta$ and $\phi$ in Fig. \ref{fig:phototurns}B.  This 
is slightly noisy because the daughter colonies are not 
distributed evenly.  The track 
also shows an interesting balance between the bottom-heaviness of {\it Volvox} 
(due to the clustering of daughters in the posterior hemisphere), which tends to 
align the axis with the $z$-direction, and the phototactic tendency to 
align the colonial axis with the direction of the light (the $x$-axis). 

\begin{figure}[t]
\begin{center}
\includegraphics*[clip=true,width=0.9\columnwidth]{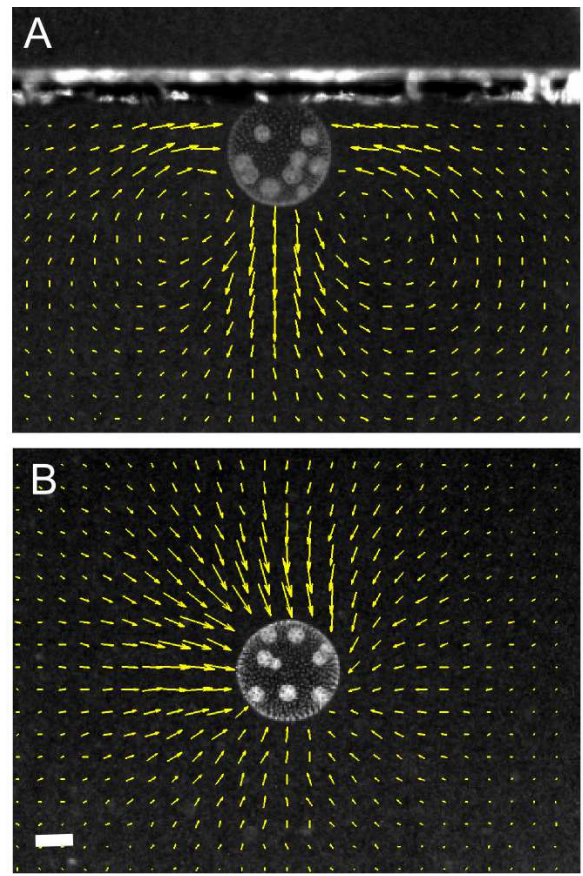}
\end{center}
\caption{\label{fig:piv} {\it Volvox carteri} swimming near a surface.  
Flow fields from particle imaging velocimetry 
of a colony swimming upward against a horizontal cover slip that 
is glued into the sample chamber (of size $2.5 \times 2.5 \times 5$ cm), as seen 
from the side (A), and the top (B). Images were taken at $\times 2$ magnification. 
Scale bar is $200$ $\mu$m.}
\end{figure}

In addition to allowing a controlled systematic disturbance to the behaviour of microorganisms 
in the form of light, this apparatus is also suitable to study the interactions of microorganisms 
with a surface, as the sample chambers can be made so large and so temperature-homogenized 
that the effect from other surfaces and thermal convection can be neglected. 
One interesting effect, discussed in more detail elsewhere \cite{waltzing}, occurs when
colonies swim upward to a horizontal surface and rotate in place.  
As the colonies are denser than the water in which they swim, and thus have a net
external force acting on them, the far-field flow around them is described 
by a stokeslet pointing downwards.  In the neighborhood of a no-slip surface, the stokeslet
induces a set of image singularities which together 
produce a characteristic lobed flow field \cite{Blake}.  This is
readily demonstrated by glueing a horizontal surface (a microscope coverslip) 
into the sample chamber and performing dual-view PIV.  The results of this are shown in 
Fig. \ref{fig:piv}A,B, illustrating the fluid flow vector fields viewed from both the side and
the top. From above, we see vectors oriented predominantly inward. 
This inward flow leads to complex behavior of nearby colonies \cite{waltzing}.

\begin{figure}[t]
\begin{center}
\includegraphics*[clip=true,width=0.95\columnwidth]{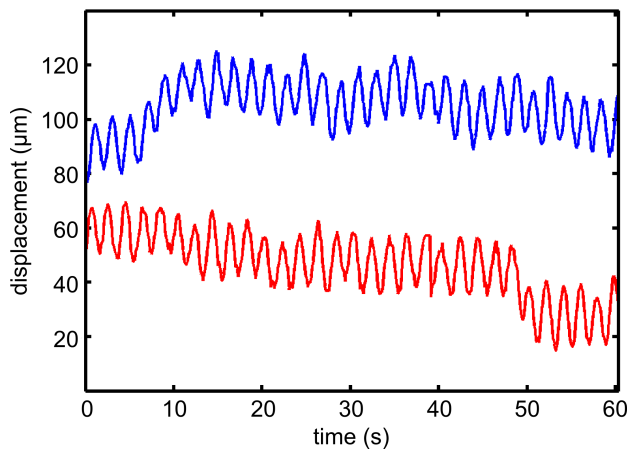}
\end{center}
\caption{\label{fig:rotation}  Tracking of spinning {\it V. carteri}.
Shown are the $x$ (red) and $y$ (blue ) coordinates
of a colony spinning near an upper surface with its
axis vertical. Recorded at $10$ fps in convection controlled chamber. The oscillations
represent a small periodic wobble in the colony centroid.}
\end{figure}

When viewed from above as in Fig. \ref{fig:piv}B
this setup also provides a means to monitor the rotational dynamics of colonies 
in great detail, providing
accurate measurements of the mean rotational frequency, the noise in rotational motion,
and lateral drifts.  As mentioned earlier, the bottom-heaviness of the colonies 
keeps the colonial axis oriented vertically, allowing the daughters 
to serve as convenient markers to track rotation.  Precise determination of time series
of rotation can then be achieved by determining the correlation between successive images
and a reference image, adjusted for centroid drift.  The centroid dynamics itself serves as
a sensitive measure of colony asymmetries, such as mismatch between the colonial axis and the
axis defined by the center of buoyancy and the geometric center.  Figure
\ref{fig:rotation} shows the two components of the centroid position for a colony rotating against
an upper surface, showing a clear periodic wobble at a frequency of $\sim 0.5$ Hz.
The systematic drifts in position, which may be due to the swimming dynamics themselves or
residual convective currents, are in any event below $2$ $\mu$m/s.

\section{Conclusions}
\label{sec:conclusions}
We presented an apparatus that can track swimming microorganisms in the size range 
$10 - 1000$ $\mu$m in 3D, without the influence of systematic bias due to beavioral 
stimuli, hydrodynamic interactions with surfaces, and convective background flows. 
As the apparatus can eliminate these biases, it can also be used to study the 
influence of each of them. The simplicity of the apparatus compared to other 3D 
tracking systems, and the software that is part of the supporting online material, 
make this system easily reproducible.

The heart of the apparatus can also be used as the basis for studies of more
complex phenomena.  For instance, the entire device can be mounted on a tiltable 
platform in order to examine the effects of varying direction of gravity with respect
to the phototactic axis.  Likewise, a rotatable chamber can be substituted for the
ones discussed here in order to examine the effects of fluid vorticity on phototactic
swimming \cite{Thorn}. 

\section{Acknowledgments}
\label{sec:acknowledgments}
We are grateful to D. Page-Croft, J. Milton, and T. Parkin for vital 
technical assistance, and to J.P. Gollub, J.T. Locsei, M. Polin and 
I. Tuval for important discussions.
This work was supported by the EPSRC, Engineering and Biological Sciences 
program of the BBSRC, the Schlumberger Chair Fund, and 
DOE grant No. DE-AC02-06CH11357.

\vfil
\eject

\end{document}